\documentclass[a4paper]{aa}
\usepackage{epsf,epsfig}

\def\correct#1 {#1}

\def\G1915{GRS $1915$+$105$}
\def\X1550{XTE J$1550$-$564$}
\def\J1655{GRO J$1655$-$40$}

\def\ie{{\em i.e. } }

\def\dd #1 {{\frac{\partial}{\partial #1}}}

\def\tom{\tilde\omega}
\def\cY{\Psi} 
\newcommand{\drond}{\partial}

\def\cs2{c_{S}^2}

\def\rhs{right-hand side }
\def\ltsima{$\; \buildrel < \over \sim \;$}
\def\simlt{\lower.5ex\hbox{\ltsima}}
\def\gtsima{$\;\buildrel>\over\sim\;$}
\def\simgt{\lower.5ex\hbox{\gtsima}}

\newcommand{\be}{\begin{eqnarray}}
\newcommand{\nonb}{\nonumber \\}
\newcommand{\ee}{\end{eqnarray}}
\def\tnabla{\tilde{\vec\nabla}}

\authorrunning{Varni\`ere and Tagger}
\titlerunning{AEI: Feeding the Corona with Alfv\'en
Waves}

\begin{document}

\title{ Accretion-Ejection Instability in magnetized disks: \\
Feeding the Corona with Alfv\'en Waves}

\author{P. Varni\`ere  and M.Tagger}

\offprints{P.~Varni\`ere \\
(pvarni@discovery.saclay.cea.fr)
}

\institute{DSM/DAPNIA/Service d'Astrophysique (CNRS URA 2052), CEA
Saclay, 91191 Gif-sur-Yvette, France}

\date{Received date; accepted date}

\abstract{We present a detailed calculation of the mechanism by which
the Accretion-Ejection Instability can extract accretion energy and
angular momentum from a magnetized disk, and redirect them to its
corona.  In a disk threaded by a poloidal magnetic field of the order of
equipartition with the gas pressure, the instability is composed of a
spiral wave (analogous to galactic ones) and a Rossby vortex.  The
mechanism detailed here describes how the vortex, twisting the footpoints of
field lines threading the disk, generates Alfv\'en waves propagating to
the corona.  We find that this is a very efficient mechanism, providing
to the corona (where it could feed a jet or a wind) a substantial
fraction of the accretion energy.   }
\maketitle

\section{Introduction}
\label{sec:Intro}
MHD models have shown that jets can be very efficient to carry away the
accretion energy and angular momentum extracted by turbulence from
accretion disks (Blandford and Payne, 1982; Lovelace {\em et al.}, 1987;
Pelletier and Pudritz, 1992), if the disk is threaded by a poloidal
magnetic field.  This fits with the observation that accretion and
ejection are intimately connected in objects ranging from protostellar
disks to X-ray binaries and AGNs.  However these models, based on
self-similar analytical computations or on numerical simulations, most
often start at the upper surface of the disk.  Although more recent
works (Ferreira and Pelletier, 1993a, 1993b, 1995; Casse and Ferreira,
2000) find solutions connecting continuously the disk and the jet, these
solutions are heavily constrained by conflicting requirements.  These
can be traced, in good part, to the fact that disk models, whether they
rely on specific instability mechanisms or on the assumption of a
turbulent viscosity, imply that the accretion energy and angular
momentum are transported {\em radially outward}.  They must thus somehow
be redirected {\em upward} to feed the jet.

The Accretion-Ejection Instability (AEI) of magnetized accretion disks,
presented by Tagger and Pellat (1999, hereafter TP99), could provide a
solution to this difficulty.  It occurs in the inner region of the disk,
in the configuration assumed by the MHD models of jets (\ie a disk
threaded by a poloidal field of the order of equipartition with the gas
thermal pressure), and it has the unique property that energy and
momentum extracted from the disk can be emitted {\em vertically} as
Alfv\'en waves propagating along magnetic field lines to the corona of
the disk.  Thus they could provide a source for a jet or a wind formed
in the corona.

\correct{This ability to emit the energy and angular momentum as Alfven wave}
was recognized in TP99, and indeed justified the name given to the
instability.  However this possibility 
was shown only in a WKB approximation, valid away from the region (the
corotation radius, where the wave rotates at the same velocity as the
gas) where Alfv\'en wave emission is most efficient.  The WKB result was
found divergent at corotation, providing a good indication that this
mechanism of vertical emission could be quite efficient.

The goal of this paper is to present a more general derivation, valid in
the corotation region.  We use a description of the waves in three
dimensions (whereas TP99 was basically a model averaged over the disk
thickness).  We are thus able to give an explicit computation of the
Alfv\'en wave emission mechanism and of its efficiency.  The main
unknown to be solved for at this stage is the fraction of the accretion
energy and angular momentum, extracted from the disk by the instability,
which will end up emitted to the corona.

The result we present is quite limited: the constraints of giving an
analytical derivation force us to use a very simplified magnetic field
geometry\correct{, namely an initially constant and vertical field,} 
and a more realistic one would certainly affect the result. 
\correct{Appendix \ref{an:Br} is dedicated to the case of a radially varying 
$B_z$ field. We show that the present computation may be apply in the case
of a slowly variable $B$ field.}
On the other hand, the result we obtain is interesting in itself: we
will show that, in linear theory, the flux of the Alfv\'en waves is
again divergent at corotation.  Although we discuss how it can be
regularized, our interpretation will be that the efficiency of the
mechanism is indeed quite high, but that we are reaching the limits of
linear theory and that the true result will most certainly be determined
by self-consistent non-linear effects.  The present work should thus be
viewed as an exploration of the complex physics involved and of its
potential efficiency, which will then have to be treated by non-linear
simulations.
\begin{figure*}[htbp]
\centering
\epsfig{file=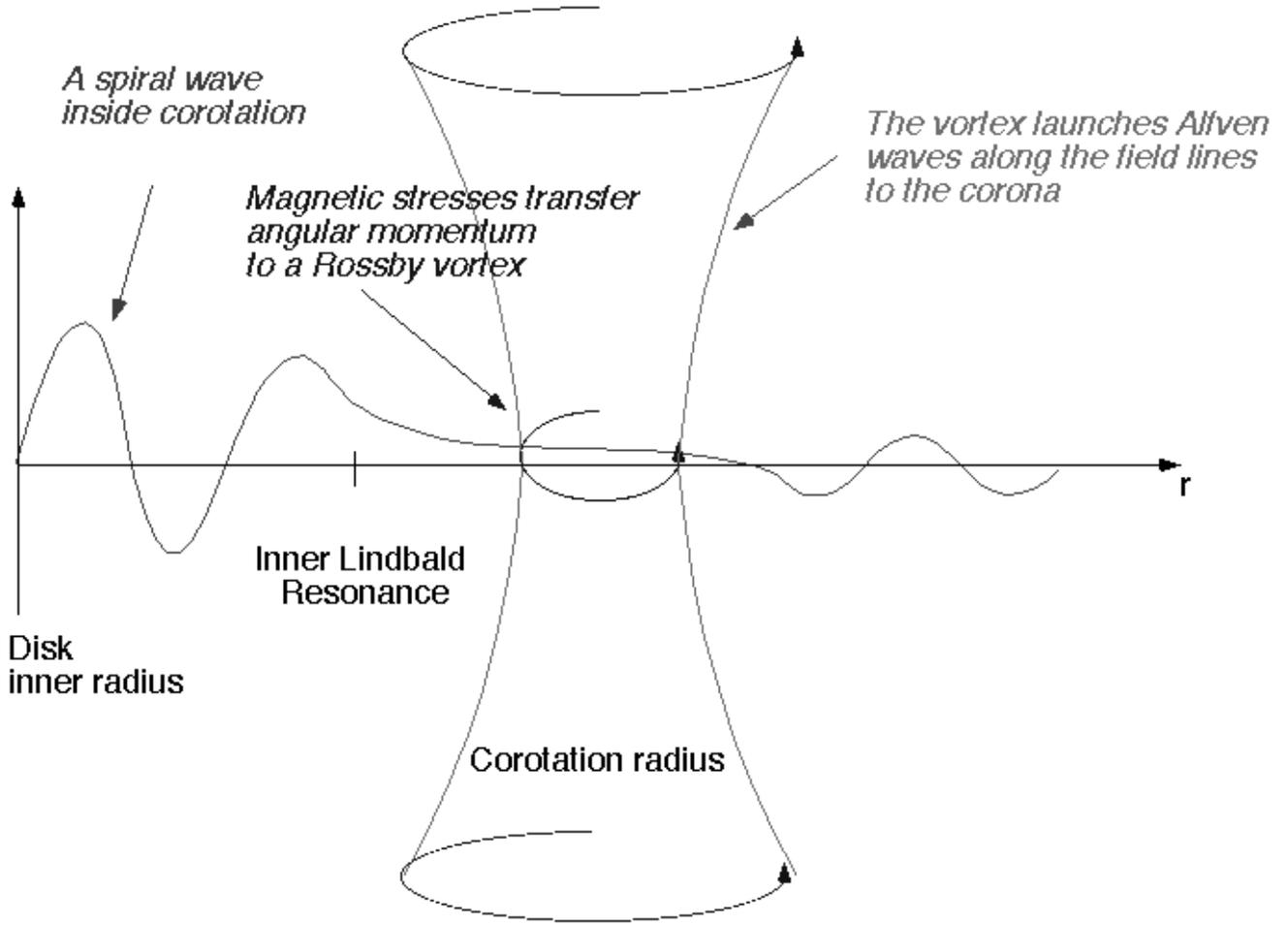,width=\textwidth}
\caption{The structure of the instability is shown here schematically as
a function of radius.  It is formed of a standing spiral density wave in
the inner part of the disk, coupled to a Rossby vortex it excites at its
corotation radius.  The spiral grows by extracting energy and angular
momentum from the disk, and depositing them in the Rossby vortex; the
latter in turn generates Alfv\'en waves propagating toward the corona of
the disk.}
\label{fig:cavite}
\end{figure*}
The paper is structured as follows: in the next section we will briefly
review the main properties of the instability, and its interest to
explain the low-frequency Quasi-Periodic Oscillation (QPO) of X-ray
binaries.  Section \ref{sec:par} will present the system of equations to
be solved, and section \ref{sec:sys} their combination into a
variational form containing the physics of the problem.  In section
\ref{sec:kz} we will compute the Alfv\'en wave flux, and we will discuss
the significance of this result in section \ref{sec:conc}.
\correct{In appendix \ref{an:Br} we will present the variational form
we obtain in a more general geometry with a radially varying $B_z$ and
the limitation it implies.}
\section{The Accretion-Ejection Instability}
\label{sec:AEI}
We will give here a short review of the main characteristics of the AEI,
and refer the interested reader to TP99, or to Varni\`ere {\em et al.}
(2002) and Rodriguez {\em et al.} (2002) where a detailed comparison
with the low-frequency QPO of black-hole binaries is given. Non-linear MHD 
simulations were performed by Caunt and Tagger (2001). The AEI is
essentially a spiral instability, similar to galactic ones but driven by
magnetic stresses rather than self-gravity.  It affects the inner region
of an accretion disk threaded by a poloidal magnetic field of the order
of equipartition with the thermal pressure of the gas ($\beta=8\pi
p/B^2$ of the order of unity), {\em i.e.} the configuration and
magnitude used in most MHD jet models.

The instability is composed of a spiral density wave and a Rossby wave
it generates at its corotation radius (the radius where the wave rotates
at the same velocity as the gas).  The spiral forms a standing pattern
between the inner edge of the disk and the corotation radius.  Because
of differential rotation, it couples to the Rossby wave, {\em i.e.} the
spiral and Rossby waves exchange energy and angular momentum in the
corotation region where, because of differential rotation they loose
their separate identities.

In this process the Rossby wave also forms a standing vortex, and the
spiral grows by storing in it the energy and angular momentum it
extracts from the inner region of the disk (thus causing accretion).

In a thin disk in vacuum, the process stops there.
\correct{This is the case in galaxies and recently Fridman {\em et al.} (2002)
observe and analyse such a vortex in NGC $157$.} 
However one must
remember that the Rossby vortex represents a torsion of the footpoints
of the field lines threading the disk.  Thus if the disk is covered by a
low-density corona, this torsion will propagate vertically along the
field lines as Alfv\'en waves: these will thus in turn transport a
fraction of the accretion energy and angular momentum to the corona,
where they could provide a source for a jet or an outflow.  The whole
process is shown schematically in figure \ref{fig:cavite}.
\section{Linearized equations}
\label{sec:par}
\subsection{Unperturbed equilibrium}

As in previous works, we consider the very simple setup  of an
infinitely thin disk threaded by a 
 vertical magnetic field \correct{adding this time the hypothesis that it is 
radially constant}.  Here
however the disk is embedded in a low density corona.  This simplified
geometry will be enough to fully characterize the emission of Alfv\'en waves
in the corona.  The equilibrium field, $\vec{B^0}=B_0\, \vec{e}_z$, is
assumed to be of the order of equipartition with the gas pressure
(plasma $\beta=8\pi p/B_{0}^2\sim 1$) in the disk.  \correct{The case of a radially
varying $B_0$ will be studied in appendix \ref{an:Br}.} 
\newline

\correct{We choose to present here results with a constant $B^0$ because
assuming that $B^{0}$ depends on $r$ creates at
equilibrium a radial magnetic pressure force which can easily become
dominant in the corona, since the other forces (gravity, Coriolis,
pressure) act on a very low density.  As a result a realistic
equilibrium will in general include a flow along the field lines, and
lead to a complex configuration limiting our ability to extract
analytical results unless artificial assumptions are made.}

\correct{On the
other hand instability requires, as found in TP99:}
\[\frac{\partial}{\partial s}\left(\frac{W\Sigma}{B_{0}^2}\right)>0\]
\correct{where $W=\kappa^2/2\Omega$ and $\kappa=\Omega\sim r^{-3/2}$ 
in a Keplerian disk.  
Thus, for this quantity to grow with $r$,
our equilibrium model needs to have a disk surface density {\em growing}
with $r$, fast enough for this condition to be fulfilled.  This appears
as an {\em ad hoc} model which does not claim to represent a real 
disk\footnote{It is worth noticing
(J. Ferreira, private communication) that the self-similar assumption
used in MHD models of jets, following Blandford and Payne, {\em always}
results in this derivative being positive and equal to $+1/2$, as a
result of the self-similarity laws.  Thus all these models are unstable
to the AEI.}.
We emphasize however that it is {\em physically consistent}, so that our
computation of the Alfv\'en wave flux remains exact and will allow us to
explore physics which applies to more general disk models.}

\subsection{Pertubations}

We work in cylindrical coordinates $[(s=\ln r), \theta, z]$.
We consider linear perturbations, described by Lagrangian displacements
$\vec \xi$ with: \be \vec V = \frac{d}{dt} \vec\xi + (\vec\xi\cdot
\vec\nabla)\vec V_0 \nonumber \ee where $\vec V$ is the perturbed
velocity, $\vec V_{0}=r\Omega(r)$ is the equilibrium rotation
velocity, and
\[\frac{d}{dt}=\frac{\partial}{\partial t}+\vec
V_{0}\cdot\vec\nabla\]
In TP99 the instability was studied by giving a full solution for its
spatial structure in cylindrical geometry; the emission of Alfv\'en
waves was then computed perturbatively, far from the corotation radius,
in the WKB limit.  The waves were described by their compressional and
torsional components,
\[D=\vec\nabla_\perp \cdot V_\perp \hskip 1 cm {\rm and} \hskip 1 cm
R=\vec\nabla_\perp \times V_\perp\]
where the subscript $\perp$ describes the components of a vector in the
plane of the disk.  We find this description inconvenient here and
rather write the displacement as:
\[B_{0}(r)\vec\xi_{\perp}=-\vec\nabla_{\perp}\Phi-\vec
e_{z}\times\vec\nabla_{\perp}\Psi\]
where again $\Phi$ and $\Psi$ represent the compressional and torsional
parts of the displacement.
\newline
 From this we write the induction equation,
\[\drond_t\vec B
+\vec\nabla\times(\vec V\times\vec B)=0:\]
giving the perturbed magnetic field as:
\begin{equation}
\vec B_\perp^1 = (\vec B^0\cdot \vec\nabla)\,\vec\xi_\perp=
B_0\, \drond_z \vec\xi_\perp \nonumber
\label{eq:bperp}\end{equation}
and from $\vec\nabla\cdot B_{1}=0$:
\begin{equation}
B_{z}^1=-\vec\nabla_{\perp}\cdot(B_{0}\,\vec\xi_{\perp})
\label{eq:bz}\end{equation}
Comparing with the solution of the continuity equation, which gives the
perturbed density $\rho=-\vec\nabla_{\perp}\cdot(\rho_{0}\,\vec\xi_{\perp})$,
one notes that equation \ref{eq:bz} is an equation of conservation for
the vertical magnetic flux.

We consider
perturbations varying as exp$[i(m\theta-\omega t)]$ (so that $m$ will be
the number of arms of the spiral).  We define a de-dimensionalized
perpendicular gradient,
\[\vec\tnabla_\perp X= r
\vec\nabla_\perp X = \drond_s X \vec e_r +imX \vec e_\vartheta.\]
We finally get the perturbed magnetic field:
\be
\vec B^1 = \frac{1}{r} \left(
\begin{array}{c}
-\drond_{z,s}\Phi  +im \drond_z \cY\\
-im\drond_z\Phi -\drond_{z,s} \cY\\
\frac{1}{r} \vec\tnabla_\perp^2\Phi
\end{array}
\right) \label{eq:B} \ee

As in TP99, our goal is to derive from the MHD equations a variational
form.  Its real (self-adjoint) part represents the energy of the
perturbations and describes their structure in the disk.  Amplification
and damping appear through imaginary terms, representing a flux of
energy: either at the boundaries of the system (as outgoing waves) or at
the corotation resonance.  The classical physics of spiral waves in
disks with differential rotation is that they have positive energy
beyond the corotation resonance ($r_{C}$ such that
$m\Omega(r_{C})=\omega$), and negative energy inside it: at $r<r_{C}$
the gas rotates faster than the wave, whose presence {\em decreases} the
total energy of the system (it releases more gravitational energy than
it costs kinetic energy).  \\
The spiral can thus be amplified if positive energy is emitted beyond
corotation (as a spiral wave emitted outward: this is the Swing
mechanism, responsible for the amplification of galactic spirals), or
stored in a Rossby vortex at the corotation radius: in TP99 this was
shown to result, in a magnetized disk, in the Accretion-Ejection
Instability.  This corotation resonance introduces, in the variational
form of TP99, a pole (a denominator proportional to
$[\omega-m\Omega(r)]$, singular at corotation).  This pole contributes,
following the Landau prescription familiar in plasma physics, an
imaginary term which represents the energy exchanged with the Rossby
vortex.  \\

In TP99 it was also found that, when one takes into account a
low-density corona above the disk, Alfv\'en waves are emitted toward the
corona; their contribution was computed in a WKB approximation, valid
only away from corotation.  This had of course a limited interest, since
we expect the Alfv\'en waves to be strongest at corotation where the
torsional motion associated with the vortex is strongest.  Here we will
present a more general formulation, showing how the energy flux of the
Alfv\'en waves appears as an additional imaginary term in the
variational form, generated at the vertical boundaries of the disk.
This will allow us to compute explicitly the Alfv\'en wave flux at
corotation.
\newline

In terms of the Lagrangian displacement, the linearized Euler equation
for a perturbation varying as $e^{i(m\vartheta -\omega t)}$ can be
written as:
\begin{eqnarray}
     (-\tom^2 +2\Omega\Omega^\prime)\ \xi_r +2i\tom\Omega\ \xi_\vartheta &=&
f_r\label{eq:Euler_r}\\
-\tom^2 \xi_\vartheta -2i\tom\Omega\ \xi_r &=& f_\vartheta
\label{eq:Euler_theta} \\
-\tom^2 \xi_z &=& f_z\label{eq:Euler_z}
\end{eqnarray}
where $\vec f=\vec F/\rho_{0}$, $\vec F$ is the force acting on the
system, $\rho_{0}(s,z)$ is the equilibrium density,
$\tom(s)=\omega-m\Omega(s)$, and the prime denotes differentiation with
respect to $s$.  In terms of $\Phi$ and $\cY$ the $r$ and $\vartheta$
components of these equations become:
\begin{eqnarray}
(\tom^2 -2\Omega\Omega^\prime)\,\drond_s \Phi + 2m\tom\Omega\Phi
\ \ \ \ \ \ \ \ \ \ \ \ \ \ \  \ \ &&\nonumber\\
-2i\tom\Omega\drond_s \cY -im(\tom^2 -2\Omega\Omega^\prime)\, \cY
&=& rB_0 f_r\label{eq:f_r}\\
2i\tom\Omega\drond_s \Phi +im\tom^2 \Phi +\tom^2 \drond_s \cY
+2m\tom\Omega\cY &=& rB_0 f_\vartheta\label{eq:f_theta}
\end{eqnarray}
For simplicity we will neglect the pressure stress, since it was found
in TP99 to be negligible, compared to the magnetic ones, in the physics
of the instability.  It could be added without change on the emission of
Alfv\'en waves.  On the other hand, this means that we do not consider
here the slow magnetosonic wave which could also propagate upward from
the disk.  We have two reasons for that: the first one is that this
would lead to a very complex problem, which goes beyond the goals of the
present paper; it would force us to consider the detailed physics at the
slow magnetosonic point above the disk surface (see {\em e.g.} Ferreira
and Pelletier, 1995, and references therein), where the gas is first
extracted from the disk.  This means that, in stationary MHD models of
jets, the physics of the slow magnetosonic wave concerns the mass
loading of the field lines, rather than the acceleration of the jet
which is more associated with the Alfv\'en wave.  Our second reason is
that we will show below how the Alfv\'en wave is mainly excited in a
resonant layer at the corotation radius, by the singular Rossby vortex
generated in the disk.  No such singularity concerns the slow wave,
which should thus not be so important for us.  We thus postpone its
consideration to future work.  In the same manner, consistent with the
thin disk approximation, we will neglect the vertical component of the
Euler equation.\\

Linearizing the contribution of magnetic stresses,
$(\vec j\times \vec B)/\rho$, we find :
\be f= \frac{\vec j^1 \times \vec B^0}{\rho_0} \label{eq:force}
\ee
where the subscripts $0$ and $1$ note equilibrium and perturbed values.
We thus get the expression of the magnetic stresses:
\be B_0 f_r &=&
V_A^2 (\drond_z B_r^1 -\drond_r B_z^1) \nonb 
B_0 f_\vartheta &=& V_A^2 (\drond_z B_\vartheta^1 -\drond_\vartheta B_z^1)
\nonumber
\ee
(where $V^2_{A}(z)=B_{0}^2/4\pi\rho_{0}$); we express this in terms of
$\Phi$ and $\cY$, and get from equations
(\ref{eq:Euler_r}-\ref{eq:Euler_theta}):
\be
&&rB_0\left[(-\tom^2 +2\Omega\Omega^\prime)\ \xi_r +2i\tom\Omega\
  \xi_\vartheta\right] =\nonumber\\
&&\ \ \ \ \ \ \ \ \ \ \ \ \ \ \ \ \ \ \ \ \ \ \ \ \ \ \ \ \ \ \ \ \ 
  -V_A^2 (\drond_s \vec\nabla^2\Phi -im \drond_{z^2} \cY)\label{eq:f_r2} \\
 \nonumber\\
&&rB_0\left[-\tom^2 \xi_\vartheta -2i\tom\Omega\ \xi_r\right]=
-V_A^2 (\drond_{s,z^2}\cY +im \vec\nabla^2\Phi)
\label{eq:f_theta2}
\ee
%
\section{The system}
\label{sec:sys}
In order to obtain the new variational form, we first write a quadratic
form built from the divergence and the curl of $\lambda B_0$ times the
Euler equation, leaving $\lambda$ unspecified for the moment.
\newline
We first apply the operator $r^2 \vec\tnabla_\perp \cdot \lambda B_0$ to
the Euler equations, {\em i.e.} we compute:
\[\frac{\partial}{\partial s}\bigl(r\lambda
B_{0}[\ref{eq:f_r2}]\bigr)+imr\lambda
B_{0}[\ref{eq:f_theta2}]\]
we get:
\be
&&\vec\tnabla_\perp
(\lambda \tom^2 \vec\tnabla_\perp \Phi)
-2\drond_s(\lambda \Omega\Omega^\prime \drond_s \Phi) \nonb
+&&2m\drond_s
(\lambda \Omega\tom)
\Phi
- 2i\vec\tnabla_\perp(\lambda
\tom\Omega\vec\tnabla_\perp\cY)\nonb
-&&im\drond_s(\lambda\tom^2) \cY +2im\drond_s(\lambda
  \Omega\Omega^\prime \cY) \nonb
=&&-\lambda V_A^2 \vec\tnabla_\perp^2 \vec\nabla^2 \Phi \nonb
&&- (\lambda \drond_s V_A^2 + V_A^2 \drond_s \lambda) (im\drond_{z^2}\cY
-\drond_s \nabla^2\Phi)\label{eq:phi_l}
\ee
Then applying the operator $r^2\vec\tnabla_\perp\times \lambda B_0$ to
the Euler equations:
\be
&& \vec\tnabla_\perp
(\lambda \tom^2 \vec\tnabla_\perp \cY)
+2m\tom\drond_s(\lambda\Omega) \cY \nonb
+&& 2i\vec\tnabla_\perp(\lambda \tom
\Omega\vec\tnabla_\perp\Phi)
+im\drond_s(\lambda\tom^2)\Phi \nonb
+&&2im\lambda\Omega\Omega^\prime
\drond_s\Phi \nonb
=&&-\lambda V_A^2\drond_{z^2}\vec\tnabla_\perp^2\cY\nonb
&&- (\lambda \drond_s V_A^2 + V_A^2 \drond_s \lambda) (im\drond_{z^2}\cY
-\drond_s \nabla^2\Phi)\label{eq:cY_l}
\ee
In order to get a variational form from these equations we need to get
rid of the term in $(im\drond_{z^2}\cY -\drond_s \nabla^2\Phi)$, present
in both.  This will be the case if $\lambda \drond_s V_A^2 + V_A^2
\drond_s \lambda =0$.  We thus make the obvious choice
$\lambda=V_{A}^{-2}(s,z)$, and equations (\ref{eq:phi_l}-\ref{eq:cY_l})
then become:
\be
  &&\vec\tnabla_\perp \left(\frac{\tom^2}{V^2_{A}} \vec\tnabla_\perp \Phi\right)
-2\drond_s\left(\frac{\Omega\Omega^\prime}{V^2_{A}}  \drond_s \Phi\right) \nonb
&&+2m\drond_s \left(\frac{\Omega\tom}{V^2_{A}}\right) \Phi
- 2i\vec\tnabla_\perp\left(
\frac{\Omega\tom}{V^2_{A}}\vec\tnabla_\perp\cY\right)\nonb
&&-im\drond_s\left(\frac{\tom^2}{V^2_{A}}\right) \cY
+2im\drond_s\left(\frac{\Omega\Omega^\prime}{V^2_{A}} \cY\right)
\nonb
&=&- \vec\tnabla_\perp^2 \vec\nabla^2 \Phi \label{eq:Phi}
\ee
and:
\be
&&\vec\tnabla_\perp
\left(\frac{\tom^2}{V^2_{A}} \vec\tnabla_\perp \cY\right)
+2m\tom\drond_s\left(\frac{\Omega}{V^2_{A}}\right) \cY \nonb
&&+ 2i\vec\tnabla_\perp\left(\frac{\tom\Omega}{V^2_{A}}
\vec\tnabla_\perp\Phi\right)
+im\drond_s\left(\frac{\tom^2}{V^2_{A}}\right)\Phi \nonb
&&+2im\frac{\Omega\Omega^\prime}{V^2_{A}}
\drond_s\Phi\nonb
&=& -\drond_{z^2}\vec\tnabla_\perp^2\cY\label{eq:Psi} \ee
We will later put these equations in a more compact form, but use for
the moment the present one which is best adapted to a variational
formulation.
%
\subsection{The Variational Form}
\label{subsec:var_form}

In order to obtain the variational form we will first integrate these
equations vertically. Assuming that the disk is covered by a
low-density corona, we integrate between $z=-z_{maz}$ and $+z_{max}$,
chosen well into the corona so that at that height only the 
\correct{Alfv\'en\footnote{As shown in Tagger {\em et al.} 1990  
the fast magnetosonic wave does not propagate in the corotation region.}
waves propagate and their vertical propagation can be described in the WKB 
approximation.}
We write the integral:
\be
{F} &\equiv& \int_{s_{min}}^{s_{max}}
\int_{z_{min}}^{z_{max}}\ \Phi^\star \ [\ref{eq:Phi}]+\cY \
[\ref{eq:Psi}]^\star \ dz ds \nonb
&=&0
\ee
where $[\ref{eq:Psi}]^\star$ is the complex conjugate of
equation~(\ref{eq:Psi}), and the radial boundaries $s_{min}$ and
$s_{max}$ will be discussed later.  We will find that, thanks to our
choice of $\lambda$, this form has the properties we wished: it is
composed of a main part, which is variational, and additional terms
which can be treated perturbatively to give amplification or damping of
the waves.\\

After some algebra, integrating by parts and grouping
terms we get:
\be
\int_{s_{min}}^{s_{max}}\int_{z_{min}}^{z_{max}}
  &\bigg\{&\frac{\tom^2 }{V^2_{A}}\
({|\vec\tnabla_\perp \cY|^2} -{|\vec\tnabla_\perp
\Phi|^2}) \nonb
&+& 2 \frac{\Omega\Omega^\prime }{V^2_{A}} \ {|\drond_s\Phi|^2} \nonb
&+& 2m \
\drond_s\left(\frac{\tom\Omega }{V^2_{A}}\right) {|\Phi|^2} \nonb
&-& 2m \tom \ \drond_s\left(\frac{\Omega }{V^2_{A}}\right) \ {|\cY|^2} \nonb
&-& \Phi^\star \vec\tnabla_\perp^2 \vec\nabla^2\Phi \ \bigg\} ds\ dz\nonb
=-\bigg[\int_{z_{min}}^{z_{max}}\frac{dz }{V^2_{A}}&\bigg(&\tom^2
\ {\Phi^\star\vec\tnabla_\perp\Phi} -2
\Omega\Omega^\prime\ {\Phi^\star\drond_s\Phi}
\nonb
&-&{\tom^2}\ {\cY\vec\tnabla_\perp\cY^\star}
+ 2im{\Omega\Omega^\prime}\  {\Phi^\star\cY}
\nonb
&-&2i {\tom\Omega}\ \left(\ {\cY\vec\tnabla_\perp\Phi^\star}
+{\Phi^\star\vec\tnabla_\perp\cY}\ \right)\bigg)\bigg]_{s_{min}}^{s_{max}}
\nonb
+{\int}_{s_{min}}^{s_{max}}
\ \bigg[\cY^\star &\drond_z& \vec\tnabla_\perp^2 
\cY\bigg]_{-z_{max}}^{z_{max}} ds
\label{eq:form}
\ee
\\

Equation (\ref{eq:form}) is equivalent to the variational form derived
in TP99, but the vertical integration (rather than the approximation of
an infinitely thin disk) will give us access to the emission of Alfv\'en
waves.  The first five terms of equation(\ref{eq:form}) appear to be
obviously hermitian, although they hide the imaginary contribution from
the corotation resonance ({\em i.e.} the growth or damping term
corresponding to the energy exchanged with the Rossby vortex) which will
be discussed in the next subsection. \\
The right-hand side is formed of boundary terms, obtained in the
integrations by parts.  All these terms are easily shown to be imaginary
({\em i.e.} correspond to growth or damping), if the boundaries are far
enough that the radial and vertical derivatives can be estimated in a
WKB approximation, $\partial_{s}=ik_{s}, \ \partial_{z}=ik_{z}$, and if
waves do propagate at the boundaries, {\em i.e.} if the wavevectors are
real so that waves can effectively transport energy away.\\
The first bracket corresponds to the flux of the wave at the radial
boundaries; as explained in TP99, a wide range of boundary conditions at
the inner disk edge allow the waves to be reflected without loss of
energy, {\em i.e.} this term vanishes or remains real at $s_{min}$.  At
$s_{max}$ it gives the flux of an outgoing wave, responsible for the
usual Swing amplification of spiral waves (driven by self-gravity in
galaxies, by pressure in the Papaloizou-Pringle instability (Papaloizou and Pringle, 1985), or by
magnetic stresses in Tagger {\em et al.} 1990).  \\
The last term is new and corresponds to a flux at the vertical
boundaries, {\em i.e.} the flux of the Alfv\'en waves emitted vertically.
This is confirmed by the fact that this term is associated with the
torsional ($\cY$) component of the perturbations.

\subsection{Corotation Resonance}
\label{subseq:corot}
In our variational form, equation (\ref{eq:form}), the corotation
resonance does not appear explicitly as it did in the equivalent form of
TP99: we do not get denominators containing $\tom$, vanishing at
corotation.  But the corotation resonance must of course be present,
since the physics described here is more general than in TP99: it is
hidden here in the singularity of $\cY$: equation (\ref{eq:Psi}) has a
singular point at corotation (where $\tom$ vanishes) because the
highest-order radial derivative of $\cY$ is proportional to $\tom^2$
(assuming, as will be checked a posteriori, that the vertical derivative
vanishes at corotation), while the terms in $\Phi$ are proportional to
$\tom$.  We thus turn, in the vicinity of corotation, to a Frob\'enius
expansion of the form:
\be
\cY& =&
a_{-1}\tom^{-1} +a_n\tom^n + b_{-1}\tom^{-1} \ln \tom +b_n \tom^n\ln
\tom\  \label{eq:F_cY} \\
\Phi& =&
c_{0}+ c_{n}\tom^n+ d_{1}\tom\ln \tom +d_n \tom^n\ln
\tom\  \label{eq:F_Phi}
\ee
where the coefficients $a,\ b,\ c,\ d$ depend on $z$.  We first note
that, in equations (\ref{eq:Phi}-\ref{eq:Psi}), the force terms (in the
right-hand sides) are small in the disk, since they are of order
$k^2V^2_{A}/\Omega^2$ (where $k$ is a radial or vertical scale length)
compared to the other ones.  Since we consider that
$\beta=2c_{S}^2/V^2_{A}\sim 1$, and $c_{s}\sim h\Omega/r$ where $h$ is
the disk thickness, these terms are small near corotation (they play of
course their role in the radial propagation of the waves, far from
corotation, see TP99).  The vertical wavenumbers will be discussed in
more details in section \ref{subsec:disp} below.  We re-arrange
equations (\ref{eq:Phi}-\ref{eq:Psi}) as:
\be
&&\vec\tnabla_\perp \left(\frac{\tom^2-2\Omega\Omega^\prime}{V^2_{A}}
  \ \vec\tnabla_\perp \Phi\right) \nonb
&& +2m\tom\drond_s \left(\frac{\Omega}{V^2_{A}}\right) \Phi
-4m^2\frac{\Omega\Omega'}{V^2_{A}}\,\Phi
\nonb
&&- 2i\vec\tnabla_\perp\left(
\frac{\Omega}{V^2_{A}}\ \vec\tnabla_\perp(\tom\cY)\right)
-im\drond_s\left(\frac{\tom^2}{V^2_{A}}\right) \cY
\nonb
&&=- \vec\tnabla_\perp^2 \vec\nabla^2 \Phi\label{eq:Phi1}
\ee
and:
\be
&&\vec\tnabla_\perp
\left(\frac{\tom^2}{V^2_{A}} \vec\tnabla_\perp \cY\right)
+2m\tom\drond_s\left(\frac{\Omega}{V^2_{A}}\right) \cY \nonb
&&+ 2i\tom\vec\tnabla_\perp\left(\frac{\Omega}{V^2_{A}}
\vec\tnabla_\perp\Phi\right)
+im\drond_s\left(\frac{\tom^2}{V^2_{A}}\right)\Phi \nonb
&&= -\drond_{z^2}\vec\tnabla_\perp^2\cY\label{eq:Psi1}
\ee
and neglect in the disk the force terms in the right-hand sides.  To
leading order ($\tom^{-1}$) equation (\ref{eq:Phi1}) gives:
\be
d_{1}=-\frac{i}{\Omega'}b_{0}
\ee
while to order $\tom^0$ equation (\ref{eq:Psi1}) gives:
\be
a_{-1}\drond_{s}\left(\frac{W}{V^2_{A}}\right)
+mb_{0}\,\frac{\Omega'^2}{V^2_{A}}
+2imd_{1}\,\frac{\Omega\Omega'^2}{V^2_{A}}=0
\ee
so that
\be
b_{0}=-\frac{a_{-1}}{m\Omega'}\,\drond_{s}\ln\left(\frac{W}{V^2_{A}}\right)
\label{eq:b0}\ee
and the following coefficients can be derived from the next
orders.

The manner in which $\Phi$ and $\cY$ project on the solutions which are
regular and singular at corotation depends on the global solution, which
must be obtained numerically as in TP99, of the problem with its
boundary conditions.  Here we will only use the fact that $\cY$ has a
singular contribution (whose Frob\'enius expansion starts with
$a_{-1}\tom^{-1}$) which makes the quadratic form, equation
(\ref{eq:form}), non variational at corotation.  Combining the singular
terms, we find their contribution to the integral:
\be
F_{Corot.}=\int ds\ \int dz
&\bigl[&\ \frac{\tom^2}{V^2_{A}}\ |\drond_{s}\cY|^2\nonb
&-&2m\tom\drond_{s}\left(\frac{\Omega}{V^2_{A}}\right)|\cY|^2\ \bigr]
\ee
Integrating by parts the first term, and retaining only the $\tom^{-1}$
term in the expansion of $\cY$, we find:
\be F_{Corot.}&=&-\int ds\int dz\ \frac{|a_{-1}|^2}{\tom^\star}\drond_s
\left(\frac{\rho_{0} W}{B_0^2}\right) \nonb
&\approx&-\int ds\ \frac{|a_{-1}|^2}{\tom^\star}\drond_s
\left(\frac{\Sigma W}{B_0^2}\right)
\label{eq:FCor}
\ee
where $\Sigma$ is the surface density of the disk.  As expected we
recover the result from TP99 that the resonant contribution to the
variational form (and hence to the growth rate of the instability) is
proportional to $\drond_s(\rho W/B_0^2)$, and will give an imaginary
contribution from the pole at corotation.  \\
Our goal now is to compute how this singular vorticity generates
Alfv\'en waves, transmitting to the corona a part of the accretion
energy and angular momentum extracted from the disk.  This flux is
readily identified, in equation (\ref{eq:form}), as the last term which
represents a contribution from the lower and upper boundaries of our
integration domain: this is thus the flux emitted vertically, and we
expect that the large amplitude and the singularity of $\cY$ will give a
strong contribution at corotation.
\section{Dispersion relation and Alfv\'en Flux}
\label{sec:kz}
\subsection{Dispersion Relation in the Corona}
\label{subsec:disp}
The flux of Alfv\'en waves to the corona appears in our variational
form, equation (\ref{eq:form}), as a surface term taken at the lower and
upper boundaries of our integration domain:
\be
F_{Alfven}={\int}_{s_{min}}^{s_{max}}
&\ &\bigg[\cY^\star\ \drond_z \vec\tnabla_\perp^2
\cY\bigg]_{-z_{max}}^{z_{max}} ds
\label{eq:Falf}\ee
We take these boundaries far enough above the disk that the density
varies smoothly, so that the vertical derivative can be computed in a
WKB approximation, $\drond_{z}\approx ik_{z}$.  A WKB approximation in
the {\em radial} direction was also used in TP99 to compute this flux. 
However, near corotation, the singularity of $\cY$ makes this radial WKB
expansion invalid.  We will show this by turning again to a Frob\'enius
expansion, but retaining the force terms in equations
(\ref{eq:Phi1}-\ref{eq:Psi1}).

Let us first consider the case of a disk in vacuum: the Alfv\'en
velocity in the corona goes to infinity, so that equations
(\ref{eq:Phi1}-\ref{eq:Psi1}) reduce to $\Delta\Phi= 0$ and
$\drond_{z^2}\cY=0$.  In a radial WKB approximation, this gives the
result familiar in spiral wave theory, that $\Phi$ varies above the disk
as $exp(-|kz|)$, where $k$ is the horizontal wavenumber, so that $\Phi$
vanishes exponentially.  On the other hand one finds that $\cY$ stays
constant with $z$.  The full vertical solution, valid across the disk, was given
by Tagger {\em et al.}, 1992.  If the coronal density is now small but
non vanishing, equation (\ref{eq:Phi1}) shows that far above the disk
$\Phi$ must be of the order of $(\Omega^2/(k^2V^2_{A\infty})\cY$, where
$k^{-1}$ is the radial scalelength of $\Phi$, and $V_{A\infty}$ is the
Alfv\'en velocity in the corona, large but not infinite.  Thus now $\Phi$ is negligible in
equation (\ref{eq:Psi1}), while the vertical derivative of $\cY$ must be
retained.  In a WKB approximation in $z$ we now expand this derivative
as:
\be
\drond_{z^2}\approx -k^2_{z}=-e_{2}\tom^2-e_{3}\tom^3+\ldots
\ee
and to lowest order of the Frob\'enius expansion
equation (\ref{eq:Psi1}) now gives $e_{2}=0$ and:
\be
a_{-1}\drond_{s}\left(\frac{W}{V^2_{A\infty}}\right)
+m\,b_{0}\,\frac{\Omega'^2}{V^2_{A\infty}}
-2e_{3\,}m\,\Omega'^2\, a_{-1}=0
\ee
Since the result obtained at the disk, equation (\ref{eq:b0}),
determines the values of $a_{-1}$ and $b_{0}$ at the base of the
corona, we can use them to get $e_3$ giving finally (assuming for simplicity that
the radial profile of $V^2_{A}$ is independent of $z$):
\be
k^2_{z}V^2_{A\infty}=\frac{\Omega\tom^3}{m\Omega'^2}\
\drond_{s}\ln\left(\frac{W}{V^2_{A}}\right)
\label{eq:kz}\ee
This result comes as a surprise, for two reasons: the first one is that,
since we consider the propagation of a perturbation which is very
localized in $s$ (since it is singular), one would have been tempted to
use a radial WKB approximation to solve equation (\ref{eq:Psi1}).  From
the first terms in the left and right-hand sides of this equation, one
would get $k^2_{z}V^2_{A\infty}=\tom^2$, {\em i.e.} the familiar
dispersion relation of Alfv\'en waves, Doppler-shifted by differential
rotation.  The answer here is simple, since with $\cY\sim\tom^{-1}$ the
first term in the right-hand side of equation (\ref{eq:Psi1}) vanishes
to lowest order in $\tom^{-1}$.  Thus the `usual' dispersion relation
applies to regular features, but not to the singular one we are
concerned with.\\

The second surprise is that $k^2_{z}$ is proportional to $\tom^3$, so
that we find $k_{z}$ real only on one side of corotation, depending on
the sign of the derivative of $W/V^2_{A}$. Thus the singular
perturbation will propagate only on one side of corotation! This can be
understood by returning to equation (\ref{eq:Psi1}) and writing it now
in terms of $Y=\tom\cY$, still neglecting the contribution of $\Phi$.
We get:
\be
\tom\vec\tnabla_\perp\frac{1}{V^2_{A}}\vec\tnabla_\perp Y
+mY\drond_{s}\left(\frac{W}{V^2_{A}}\right)
=k^2_{z}\vec\tnabla_\perp^2\left(\frac{Y}{\tom}\right)
\label{eq:Ross}\ee
Let us first neglect the magnetic tension force, on the right-hand side.
Equation (\ref{eq:Ross}) gives, in a radial WKB approximation
($\drond_{s}\approx ik$):
\be
\tom=\frac{mV^2_{A}}{k^2+m^2}\ \drond_{s}\left(\frac{W}{V^2_{A}}\right)
\ee
This is just the dispersion relation of Rossby waves, Doppler-shifted by
differential rotation and including the radial gradient of $V^2_{A}$! 
Because this dispersion relation is odd in $\tom$ (whereas more usual
ones are even) it allows Rossby waves to propagate only on one side of
corotation\footnote{For a more detailed discussion of Rossby waves with
differential rotation, see Tagger, 2001.  In particular, this shows how
a Rossby wave in a disk \emph{always} collapses to a singularity, the
one analyzed here, after a finite time, of the order of a rotation
period.}.\\

The meaning of equation (\ref{eq:Ross}) becomes simple: it describes how
Rossby waves, forming the singular part of our full solution for $\cY$,
will now propagate upward along the field lines as what we might call
Rossby-Alfv\'en waves.  If the radial derivative of $W/V^2_{A}$ is
positive (this is the instability criterion of TP99), the wave
propagates only beyond corotation and thus carries a positive energy
flux, so that its formation in the disk and propagation along the field
lines destabilize the negative energy spiral wave inside corotation, which is
the main component of our instability.\\
\correct{This makes the physics of the coupling between the instability
and Alfv\'en waves much more complex than the resonance, found by Curry
and Pudritz (1996), between normal modes of the Magneto-Rotational
Instability (MRI) and Alfv\'en waves.  The main reason is that they work
in a vertical WKB approximation, assuming that the modes have a fixed
vertical wavenumber $k_{z}$; it is worth remembering here the result of
a more complete vertical solution (Tagger {\em et al.}, 1992): the AEI
corresponds to solutions with $n_{z}=0$ nodes across the disk height,
whereas the MRI corresponds to solutions with $n_{z}\leq 1$.  For
$n_{z}\gg 1$, most unstable when $\beta\gg 1$, a vertical WKB
approximation can be used.  Keeping $k_{z}$ fixed results in finding a
resonance away from corotation, at $\tom=k_{z}V_{A}$, and thus ignoring 
the coupling with the Rossby vortex. It also ignores the difficulties, 
encountered here, associated with the vertical variation of the 
Alfv\'en velocity.}
\subsection{Alfv\'en waves Flux}
We can now compute the flux of Alfv\'en waves, appearing in the
right-hand side of equation (\ref{eq:form}):
\be
F_{Alfven}=i\int_{s_{min}}^{s_{max}} [k_z |\vec\tnabla_\perp
\cY|^2]_{z_{min}}^{z_{max}} \ ds\label{eq:flux}
\ee
where we have assumed that the upper boundary is high enough above the
disk, so that the vertical derivative can be replaced by $ik_{z}$.  This
contribution was computed in TP99, in a radial WKB approximation valid far from
corotation.  Here our goal is to compute the localized contribution of
the resonance, {\em i.e.} the flux of Alfv\'en waves launched from the
Rossby vortex, and more precisely the flux associated with the singular
part of $\cY$ as computed in the previous sections.  Thus we retain
$\cY\approx a_{-1}\tom^{-1}$ near corotation, and take into account only
the region beyond corotation where (if the radial gradient of
$W/V^2_{A}$ is positive) the Rossby-Alfv\'en wave propagates, so that
$k_{z}$ is real.  We choose for $k_{z}$ the sign of $z$, in order to
have waves propagating away at $z_{min}$ and $z_{max}$.  We get:
%
\be
F_{Alfven}\approx 2i|a_{-1}|^2&&\int_{0}^{s_{max}} ds\ 
\frac{\tom^{3/2}}{|\tom|^4}\nonb
&&\ \ \ \ \ \ \ \ \frac{m^2|\Omega'|}{V_{A\infty}}\ \biggl[\Omega
\drond_{s}\ln\left(\frac{W}{V^2_{A}}\right)\biggr]^{1/2}
\label{eq:FAlfven}
\ee
This integral is strongly divergent at corotation, as expected.  Various
effects can regularize it, in particular the presence of an imaginary
part of $\omega$, or the fact that most of our derivations fail at
radial scales of the order of the disk half-thickness $h$.  Let us consider
in particular equation (\ref{eq:Phi1}): we have solved it in the disk by
a Frob\'enius expansion, neglecting altogether the force term, on the
right-hand side, although it contains the highest-order radial
derivative. Our reason was that this term is small as long as $\tom$ is
not too small:  remembering that we assume the plasma $\beta$ to be of
the order of 1 in the disk, we have:
\be
V^2_{A}\sim c^2_{S}\sim\left(\frac{h}{r}\right)^2 \Omega^2
\ee
so that this term, acting on the divergent part of $\Phi$, is of the
order of
\be
\left(\frac{h}{r}\right)^2 \left(\frac{\Omega}{\tom}\right)^2\nonumber
\ee
compared to the one we retain. This means that our expansion breaks down
when $\tom\sim (h/r)\,\Omega$, or $s\sim h/r$, and we will take this as the
lower bound in our integral, equation (\ref{eq:FAlfven}). It is quite
possible that the full solution, at $s< h/r$, would still give a
divergent result, but then the growth rate of the instability,
estimated in TP99 to be
\be Im(\omega)\sim \left(\frac{h}{r}\right) Re(\omega)\nonumber
\ee
would give a similar regularization of the integral.  Thus we get a conservative 
estimate for the flux transported toward the corona by Alfv\'en waves:
assuming that all the gradients are of the order of 1, we get:
\be
F_{Alfven}\sim 
|a_{-1}|^2\
\left(\frac{r}{h}\right)^{3/2}
\frac{1}{\Omega V_{A\infty}}
\ee
On the other hand, the flux \emph{deposited} at corotation in the vortex
(which is the energy removed from the central region of the disk,
causing accretion) is given by equation (\ref{eq:FCor}).  As in the
classical problem of Landau damping, its imaginary part is given by a
Cauchy residue:
\be
F_{Corot.}=\frac{|a_{-1}|^2}{m|\Omega'|}\ \partial_{s}\left(\frac{\Sigma
W}{B^2_{0}}\right)
\ee
Both fluxes are related to the singularity of $\cY$ at corotation, and
thus proportional to $|a_{-1}|^2$.  Assuming again for simplicity that
$\beta\sim 1$ and that the radial gradients contribute numbers of order
unity, we thus get for their ratio a very simple expression:
\be
\frac{F_{Alfven}}{F_{Corot.}}\sim \left(\frac{\rho_{\infty}}{\rho_{D}}
\right)^{1/2}\
\left(\frac{r}{h}\right)^{3/2}
\label{eq:fluxratio}
\ee
where we have used $\Sigma\sim\rho_{D} c_{S}/\Omega$, and $\rho_{D}, \
\rho_{\infty}$ are the densities in the disk and in the corona.  The
first term in the \rhs is \emph{a priori} small.  It is typical of
magnetic braking processes, and leads to a weak efficiency of mechanisms
of magnetic origin coupling the disk to the corona.  The good surprise
for us here is that it is multiplied by the second term, which can be
quite large (typically a few hundred to a thousand in the disks of X-ray
binaries).  The flux ratio can then be a significant fraction of unity,
as soon as the corona has a density which is not vanishingly small. For higher density making our estimate larger than 1, a full 3D computation taking into account the finite thickness of the disk would be required.\\

\subsection{Non-Vertical Unperturbed Magnetic Field}

\correct{In the body of the paper we have studied the case of a
	vertical and constant unperturbed field.  In Appendix
	\ref{an:Br} we discuss the effect of a field which is still
	straight but depends on the radius.  The analytical computation
	cannot be fully completed, mainly because no simple equilibrium
	exists without a flow along the field lines; but we show that within 
	reasonable bounds (that the magnetic stress term does not exceed the 
	centrifugal and gravitation forces) our conclusions should not change.
\newline
	On the other hand a realistic model should also include the
	curved geometry of the magnetic field, as obtained in jet
	models.  In this case the situation becomes much more complex
	because in general the three basic MHD waves become coupled by
	the geometry, in addition to the coupling by differential
	rotation studied here.  In a straight field the Alfv\'en and
	slow magnetosonic wave are decoupled: this has allowed us to
	defer the consideration of slow magnetosonic waves, together
	with vertical motion, to separate work. In a curved field the Alfv\'en 
	wave will include vertical motion, and we should in principle include 
	all three components of the displacement and all three MHD waves. 
\newline	
	The mixing can still be weak if the scales are very different. 
	For instance, the characteristic wavelength of the slow wave is
	of the order of $c_{S}/\tom$, {\em i.e.} of the disk scale
	height if we are not too close to corotation.  Thus if the
	magnetic field is curved on a large scale (of the order of $r$),
	the mixing of the waves is weak and our conclusion
	should not change much. In particular the Alfv\'en wave is excited much more
	efficiently than the slow wave, for which we expect no singular
	source analogous to the Rossby vortex for the Alfv\'en wave.}

\correct{On the other hand more elaborate jet models ({\em e.g.}
	Casse \& Ferreira, 2000) show that a slow magnetosonic point forms
	above the disk.  The field lines are sharply bent in its
	vicinity.  But there the Alfven velocity is already much higher
	than the sound velocity.  This disparity should again maintain a
	separation between slow and Alfv\'en waves, and allow the latter
	to propagate the vorticity from the Rossby wave.  A realistic
	computation goes far beyond our present abilities.  But we note
	that the coupling of waves by geometric effects might introduce
	interesting new channels to deposit energy and momentum from the
	wave in the corona.}

\section{Conclusion}
\label{sec:conc}

We have presented a computation of the flux of Alfv\'en waves emitted to
the corona of a magnetized disk by the Accretion-Ejection Instability.
This means that we have justified here the name chosen by TP99: the
instability is a spiral density wave, which grows by extracting energy
and angular momentum from the disk (thus causing accretion) and
transferring them \emph{radially outward} to the Rossby vortex at
corotation; a significant fraction, given by equation
(\ref{eq:fluxratio}), of this flux is then transmitted \emph{vertically}
to the corona as Alfv\'en waves.  We expect that, if the Alfv\'en waves
can deposit their energy and momentum in the corona, this would be an
ideal mechanism to feed a wind or jet directly from the accretion
process in the disk.

The amplification of the wave (and thus the flux deposited by the spiral
in the vortex) and the flux transmitted to Alfv\'en waves are both
linked to the singularity of the vortex.  This allows us to give in a
very simple form a result of paramount importance in the physics of
accretion disks and jets: an estimate of the fraction of the accretion
energy, extracted from the inner region of the disk, which will end up
in the corona where it might feed a jet.  This fraction is of the order
of unity if the coronal density is not too low (typically $10^{-4}$ of
the density in the disk would be sufficient, in an X-ray binary).

In order to obtain analytically a physically consistent result, we have
had to use a very artificial configuration of the equilibrium magnetic
field, vertical and independent of $r$.  On the other hand this has
allowed us, proceeding rigorously by perturbation of a variational form,
to obtain an exact result clarifying the role and the physical nature
of the singularity of the Rossby vortex at corotation. We can thus
expect that these results would survive less stringent assumptions on
the equilibrium configuration.

However this must be taken carefully: our final result is in fact
divergent at the corotation radius, and regularized by the
effect of the finite thickness of the disk, or by the growth rate of the
instability.  In both cases, it depends on the density in the corona of
the disk. Thus we believe that the end result should be obtained from a
self-consistent, non-linear description where the growth of the instability
itself affects the evolution of the magnetic geometry and the mass
loading of the corona.

In this respect it is worth mentioning one of the results of stationary
MHD jets models: in these models, as the gas is accelerated along the
field lines it passes a slow magnetosonic point where the field lines
bend outwards.  Magnetocentrifugal acceleration can then proceed and
leads to the formation, higher up and further out, of an alfv\'enic
point.  The slow magnetosonic point is thus associated with the mass
loading of the field lines, and the alfv\'enic point to the
acceleration.  By analogy we can thus expect that, while the Alfv\'en
waves described in the present work allow to accelerate the gas, the
instability can also generate slow magnetosonic waves which will lift
the gas above the disk.  The coupling of the instability to the slow
wave will be the object of a forthcoming paper.

\appendix
\section{Radial dependence of $B_z$}
\label{an:Br}

	\correct{ In the body of the present article we made the simple
	assumption that the equilibrium field $\vec{B^0}$ was vertical
	and constant, allowing us to get an analytical derivation of the
	Alfven waves flux.  In this appendix we give the general
	derivation in the case where the field is still straight along
	$z$ but depends on $s$.  We will follow the same computation as
	in the main text, referring to the corresponding equations.}
\newline
		
\correct{
	The first modification appears in the contribution of magnetic stresses
	$(\vec j\times \vec B)/\rho$. We now have to take into account 
	the equilibrium current, $j^0$:}
\be f=\frac{\vec j^0 \times \vec B^1}{\rho_0}+ \frac{\vec
j^1 \times \vec B^0}{\rho_0}- (\vec j^0 \times \vec B^0)\
\frac{\rho_1}{\rho_0^2} \label{eq:force2}\ee

\correct{
	We get $\rho_{1}$ from the continuity equation,}
\be\rho_1 = - \vec\nabla
\cdot (\rho_0\vec \xi)\label{eq:rho1}
\ee
\correct{
	which gives after some transformations, and using equation (\ref{eq:bz}):}
\be
B_0 \frac{\rho^1}{\rho_0} &=& B_z^1
-\frac{1}{r^2}\,g(rB_0\xi_r)\label{eq:b1}
\ee
\correct{
	Where the function $g$ is defined by:}
\be g(rB_0\xi_r) =
\left(\left[\frac{\rho_0}{B_0}\right] -
\frac{B_0^2}{\tom^2\rho_0}[B_0]\frac{\partial^2}{\partial z^2}
\right)(rB_0\xi_r)\label{eq:g2}
\ee
\correct{
	and $rB_0\xi_r = -\drond_s \Phi +im\cY$.  Here angular brackets note the
logarithmic derivative with respect to $s$, $[X]= 1/X\ \drond_s X$, and
$g$ comes from the departure, due to magnetic pressure, from the
keplerian rotation curve. It is weak in the disk for $\beta \sim 1$; it
might become important in the corona, but this would require to take into account the 
velocity flow associated with the formation of the jet. This full computation
is beyond our present capabilities.}
\newline
\correct{
We thus get the expression of the magnetic stresses:}
\be B_0 f_r &=&
V_A^2 (\drond_z B_r^1 -\drond_r B_z^1) -B_0\frac{\drond_r B_0}{\rho_0}
B_z^1 + B_0\frac{\drond_r B_0}{\rho_0}B_0 \frac{\rho^1}{\rho_0}\nonb B_0
f_\vartheta &=& V_A^2 (\drond_z B_\vartheta^1 -\drond_\vartheta B_z^1)
\nonumber
\ee
\correct{
	Expressing this in terms of $\Phi$ and $\cY$ we get:}
\be
rB_0\left[(-\tom^2 +2\Omega\Omega^\prime)\ \xi_r +2i\tom\Omega\
  \xi_\vartheta\right] =&\nonumber\\
  -V_A^2 (\drond_s \vec\nabla^2\Phi -im \drond_{z^2} \cY)
-\frac{V_A^2}{r^2} &[B_0]\, g(rB_0\xi_r)\label{eq:f_r22} \\
& \nonumber\\
rB_0\left[-\tom^2 \xi_\vartheta -2i\tom\Omega\ \xi_r\right]=\ \ \ \ \ \ \ \
\ \ \ \ &\nonumber\\
-V_A^2 (\drond_{s,z^2}\cY &+im \vec\nabla^2\Phi)
\label{eq:f_theta22}
\ee
\correct{
	which generalize equations \ref{eq:f_r2} and \ref{eq:f_theta2}.}

\subsection{System}
	
	\correct{ From these new equations, using the same method as in
	section \ref{sec:sys} we obtain the following parametric
	system:} 
\be 
&&\vec\tnabla_\perp (\lambda \tom^2
	\vec\tnabla_\perp \Phi) -2\drond_s(\lambda \Omega\Omega^\prime
	\drond_s \Phi) \nonb
+&&2m\drond_s
(\lambda \Omega\tom)
\Phi
- 2i\vec\tnabla_\perp(\lambda
\tom\Omega\vec\tnabla_\perp\cY)\nonb
-&&im\drond_s(\lambda\tom^2) \cY +2im\drond_s(\lambda
  \Omega\Omega^\prime \cY) \nonb
=&&-\lambda V_A^2 \vec\tnabla_\perp^2 \vec\nabla^2 \Phi \nonb
&&- (\lambda \drond_s V_A^2 + V_A^2 \drond_s \lambda) (im\drond_{z^2}\cY
-\drond_s \nabla^2\Phi)\nonb
&&+(\lambda V_A^2 +\lambda \drond_s V_A^2 + V_A^2 \drond_s \lambda)
\drond_s\left(\frac{[B_0]}{r^2} g(rB_0\xi_r)\right)\label{eq:phi_l2}
\ee
\correct{
	and}
\be
&& \vec\tnabla_\perp
(\lambda \tom^2 \vec\tnabla_\perp \cY)
+2m\tom\drond_s(\lambda\Omega) \cY \nonb
+&& 2i\vec\tnabla_\perp(\lambda \tom
\Omega\vec\tnabla_\perp\Phi)
+im\drond_s(\lambda\tom^2)\Phi \nonb
+&&2im\lambda\Omega\Omega^\prime
\drond_s\Phi \nonb
=&&-\lambda V_A^2\drond_{z^2}\vec\tnabla_\perp^2\cY\nonb
&&- (\lambda \drond_s V_A^2 + V_A^2 \drond_s \lambda) (im\drond_{z^2}\cY
-\drond_s \nabla^2\Phi)\nonb
&&+(\lambda V_A^2 +\lambda \drond_s V_A^2 + V_A^2 \drond_s \lambda)
\frac{im}{r^2} [B_0]\, g(rB_0\xi_r)\label{eq:cY_l2}
\ee

\correct{ In order to get rid of the term in $(im\drond_{z^2}\cY
	-\drond_s \nabla^2\Phi)$ we make the same choice as in
	the constant-$B^0$  case, namely $\lambda=V_{A}^{-2}(s,z)$.  The
	system equivalent  to (\ref{eq:Phi}-\ref{eq:Psi}) is:}
\be
  &&\vec\tnabla_\perp \left(\frac{\tom^2}{V^2_{A}} \vec\tnabla_\perp \Phi\right)
-2\drond_s\left(\frac{\Omega\Omega^\prime}{V^2_{A}}  \drond_s \Phi\right) \nonb
&&+2m\drond_s \left(\frac{\Omega\tom}{V^2_{A}}\right) \Phi
- 2i\vec\tnabla_\perp\left(
\frac{\Omega\tom}{V^2_{A}}\vec\tnabla_\perp\cY\right)\nonb
&&-im\drond_s\left(\frac{\tom^2}{V^2_{A}}\right) \cY
+2im\drond_s\left(\frac{\Omega\Omega^\prime}{V^2_{A}} \cY\right)
\nonb
&=&- \vec\tnabla_\perp^2 \vec\nabla^2 \Phi +
\drond_s\left(\frac{[B_0]}{r^2} g(rB_0\xi_r)\right)\label{eq:Phi2}
\ee
and:
\be
&&\vec\tnabla_\perp
\left(\frac{\tom^2}{V^2_{A}} \vec\tnabla_\perp \cY\right)
+2m\tom\drond_s\left(\frac{\Omega}{V^2_{A}}\right) \cY \nonb
&&+ 2i\vec\tnabla_\perp\left(\frac{\tom\Omega}{V^2_{A}}
\vec\tnabla_\perp\Phi\right)
+im\drond_s\left(\frac{\tom^2}{V^2_{A}}\right)\Phi \nonb
&&+2im\frac{\Omega\Omega^\prime}{V^2_{A}}
\drond_s\Phi\nonb
&=& -\drond_{z^2}\vec\tnabla_\perp^2\cY+ \frac{im[B_0]}{r^2}
g(rB_0\xi_r)
\label{eq:Psi2} 
\ee

\subsection{The Variational Form}

\correct{
After some algebra, integrating by parts and grouping
terms we get the equivalent of equation (\ref{eq:form})}
\be
\int_{s_{min}}^{s_{max}}\int_{z_{min}}^{z_{max}}
  &\bigg\{&\frac{\tom^2 }{V^2_{A}}\
({|\vec\tnabla_\perp \cY|^2} -{|\vec\tnabla_\perp
\Phi|^2}) \nonb
&+& 2 \frac{\Omega\Omega^\prime }{V^2_{A}} \ {|\drond_s\Phi|^2} \nonb
&+& 2m \
\drond_s\left(\frac{\tom\Omega }{V^2_{A}}\right) {|\Phi|^2} \nonb
&-& 2m \tom \ \drond_s\left(\frac{\Omega }{V^2_{A}}\right) \ {|\cY|^2} \nonb
&-& \Phi^\star \vec\tnabla_\perp^2 \vec\nabla^2\Phi \nonb
&+& \Phi^\star \drond_s \left(
\frac{[B_0]}{r^2} \ g(rB_0\xi_r)\right) \nonb
&-& im \cY\ \frac{[B_0]}{r^2} \ g(rB_0\xi_r^\star) \ \bigg\} ds\ dz\nonb
=-\bigg[\int_{z_{min}}^{z_{max}}\frac{dz }{V^2_{A}}&\bigg(&\tom^2
\ {\Phi^\star\vec\tnabla_\perp\Phi} -2
\Omega\Omega^\prime\ {\Phi^\star\drond_s\Phi}
\nonb
&-&{\tom^2}\ {\cY\vec\tnabla_\perp\cY^\star}
+ 2im{\Omega\Omega^\prime}\  {\Phi^\star\cY}
\nonb
&-&2i {\tom\Omega}\ \left(\ {\cY\vec\tnabla_\perp\Phi^\star}
+{\Phi^\star\vec\tnabla_\perp\cY}\ \right)\bigg)\bigg]_{s_{min}}^{s_{max}}
\nonb
+{\int}_{s_{min}}^{s_{max}}
\ \bigg[\cY^\star &\drond_z& \vec\tnabla_\perp^2 
\cY\bigg]_{-z_{max}}^{z_{max}} ds
\label{eq:form2}
\ee

\correct{ The only new terms compared to equation \ref{eq:form} are the 
ones containing the function $g$. One easily checks, by expanding $g$
and $\xi_{r}$, that these terms  are also hermitian. Therefore \ref{eq:form2} 
is also a variational form. We can draw the same conclusion as in
the constant-$B^0$ case.}
\newline


\subsection{Dispersion Relation and Alfv\'en flux}

\correct{
	Taking into account the gradient of $B^0$ makes the analytical derivation
	of the dispersion relation and Alfv\'en flux much more complex and we
	will not attempt it  here.} 

	\correct{ If we make the additional assumption that $B_{0}$
	depends on $r$ only weakly, {\em i.e.} that the current $j^0$ is
	weak, we can get rid of the influence of the terms containing
	$g$ .  This assumption is equivalent to requiring that the
	magnetic stress term is at most of the order of the centrifugal
	force, which should be the case in a realistic jet model
	including an equilibrium flow along the field lines.  Using
	$V_{A\infty}$ as the Alfv\'en velocity in the corona we obtain
	the condition:}
\[\eta(s)=[B_{0}]V_{A\infty}^2\sim r^2\Omega^2\]

\correct{In this manner the conclusion presented in the body of this
article for a constant vertical $B^0$ field can be extended to a weakly
varying $B^0_z$.}

\bibliographystyle{plain}

\begin{thebibliography}{99}

\bibitem{BP82}Blandford, R. D., and Payne, D. G.,
1982, MNRAS {\bf 199}, 883%

\bibitem{CF00}
Casse, F., Ferreira, J., A\&A, 2000, {\bf 353}, 1115.%

\bibitem{Caunt}
Caunt, S. and Tagger, M., 2001, A\& A, {|bf 376}, 1095.%

\bibitem{CP96} Curry,C., Pudritz, R.E., 1996, MNRAS, {\bf 281}, 119-136.

\bibitem{FP1} Ferreira, J.; Pelletier, G., 1993a, A\&A, {\bf 276}, 625. %

\bibitem{FP2} Ferreira, J.; Pelletier, G., 1993b, A\&A, {\bf 276}, 637. %

\bibitem{FP3} Ferreira, J.; Pelletier, G., 1995, A\&A, {\bf 295}, 807. %

\bibitem{F02} Fridman, A.M.; Khoruzhii, O.V.; Lyakhovich, V.V.; Sil'chenko, O.K.;
Zasov, A.V.; Afanasiev, V.L.; Dodonov, S.N.; Boulesteix, J., 2002, 
accepted by A\&A, astro-ph/0103507. %

\bibitem{lov} Lovelace, R.V.E., Wang, J.C.L. and Sulkanen, M.E., 1987, 
ApJ {\bf 315}, 504.%

\bibitem{PP} Papaloizou, J. C. B., Pringle, J. E., 1985, MNRAS {\bf 213},
799.%

\bibitem{PP92} Pelletier, G., Pudritz, R. E.,  1992,  ApJ, {\bf 394},
 117.%

\bibitem{ROD00}Rodriguez, J., Varni\`ere, P., Tagger. M., and
Durouchoux, P., 2002, {\em to be published in} A\& A %

\bibitem{THSP90} Tagger, M.; Henriksen, R. N.; Sygnet, J. F.; Pellat, R.,
Apj, {\bf 353}, 654-657, 1990.

\bibitem{T91} Tagger, M., Pellat, R. and Coroniti, F.C., 1992, ApJ, {\bf
393},708.%

\bibitem{Tagger92}
Tagger, M., Pellat, R. and Coroniti, F., 1992, ApJ 393, 708

\bibitem{TP99} Tagger, M., and Pellat, R., 1999, A\&A, {\bf 349} 1003
(TP99)%

\bibitem{T01} Tagger, M., 2001, A\& A {\bf 380}, 750%

\bibitem{VR01} Varni\`ere, P., Rodriguez, J., Tagger, M.,
 2002, {\em to be published in} A\& A %

\end{thebibliography}

\end{document}